\documentclass[%
preprint,
 amsmath,amssymb,
 aps,
]{revtex4-1}

\usepackage{graphicx}
\usepackage{dcolumn}
\usepackage{bm}

\begin{document}

\preprint{}

\title{Robust method to determine the resolution of a superlens by analyzing the near-field image of a two-slit object}

\author{B. D. F. Casse}
 \altaffiliation[Now at: ]{Physical Sciences Inc., 20 New England Business Center, Andover, MA 01810-1077}
\author{W. T. Lu}%
\author{Y. J. Huang}
\author{S. Sridhar}
\email{S.Sridhar@neu.edu}
\affiliation{%
 Electronic Materials Research Institute and Department of Physics,\\
 Northeastern University, Boston, MA 02115.
}%

\date{\today}

\begin{abstract}
In the last decade, metamaterials-based superlenses, with a resolution below Abbe's diffraction limit, have emerged. To obtain a rough estimate of the resolution of such superlenses, imaging of two subwavelength slits, separated by a subwavelength gap \textit{d} is typically performed. The resolution $\Delta $ of the lens corresponds to the minimum possible gap $d_{min} $ for which a distinct image of the two slits can be resolved ($\Delta \sim d_{min} $). In this letter, we present a more quantitative estimate of the resolution of manufactured lenses by fitting analytical near-field image profiles, obtained from imaging a two-slit object with a theoretical negative-index lens of known resolution, to experimental data. We conclude the discussion by applying our analytical method to 3 case examples of superlensing from the literature. As shown, this method is particularly attractive for rapidly assessing the performance of fabricated superresolution lenses.
\end{abstract}

\pacs{Valid PACS appear here}
\maketitle


\section{Introduction}

In recent years, a new class of artificially engineered materials, so--called metamaterials \cite{Shalaev07, Hoffman07, Lezec07, Moser05}, has attracted considerable attention, particularly because of their potential for super-resolution imaging \cite{Pendry00}.  Following different engineering approaches, researchers have demonstrated metamaterials-based superlenses that have resolutions well below Abbe's diffraction limit \cite{Kawata09, Liu08, Fang05, Taubner06, Casse09, Smolyaninov07, Merlin07}. For the various superlenses, the resolution has been experimentally estimated by imaging two subwavelength slits at a distance \textit{d }apart (where $d<<\lambda $: wavelength). Successful recovery of the image of the slit objects with the smallest possible gap \textit{$d_{min} $} indicates the resolution $\Delta $ of the lens (i.e. $\Delta \sim d_{min} $). The drawback of this method is that it is time-consuming to determine \textit{$d_{min}$} exactly, and trial and error may be required. By using a trial and error technique, it is easy to miss \textit{$d_{min} $}, which can be exemplified from cases in the literature (discussed later in this letter).

\begin{figure}
  \includegraphics[scale=.7]{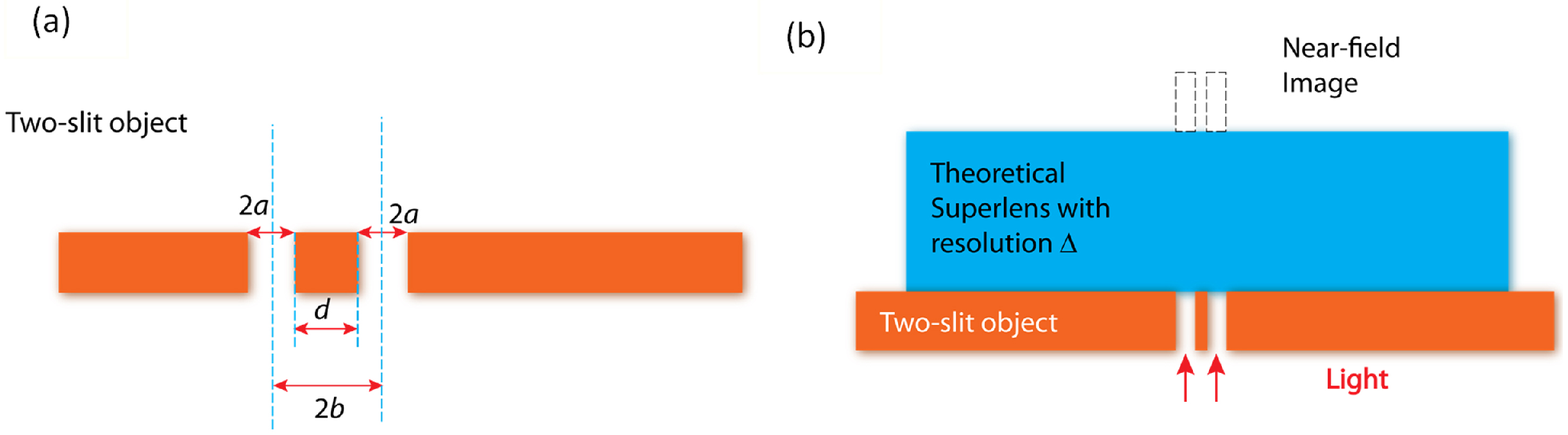}\\
  \caption{\textbf{(a)} Schematic of a two-slit object, which consists of two void slits milled in thin metallic film. The slits have identical widths $2a$ with a spacing $d$ apart. The center-to-center distance of the slits is $2b$. \textbf{(b) }Illustration of the analytical technique used to determine the resolution of a lens: The near-field profile of a two-slit object (having the same geometrical dimensions as the actual objects), imaged by a theoretical lens with resolution $\Delta$, is analytically computed. The profile of the analytical near-field image, generated by the hypothetical lens, is then fitted to the profile of the experimental near-field image data.}\label{bdfc_fig1}
\end{figure}

In this letter, we present an analytical method to obtain a more quantitative estimate of the resolution of a superlens by analyzing the near-field imaging data of a two-slit void object, as shown in Fig.~\ref{bdfc_fig1}(a). The idea is to numerically compute the near-field distribution of a two-slit object in the vicinity of a theoretical flat lens (with known resolution), shown in Fig.~ \ref{bdfc_fig1}(b), and vary the resolution of the theoretical lens, in small increments, to obtain a near-field profile which is as close as possible to the one experimentally obtained from near-field scanning optical microscope (NSOM) data. The advantage of our method is that it is possible to obtain a more accurate value for the resolution of the lens and that it is no longer necessary to find \textit{$d_{min}$} experimentally by trial and error. The method yields the correct resolution of the fabricated prototype even with data obtained with a slit gap \textit{$d>d_{min} $}. We demonstrate that the method is robust by tackling 3 case examples of superlensing from the literature and accordingly extracting the resolution of the superresolution lenses from transmitted near-field data.

\section{Analysis}

To understand the methodology, we review Pendry's theoretical ``perfect'' lens concept \cite{Pendry00}. Pendry's perfect lens  with $n=\varepsilon =\mu =-1$ has infinite spatial resolution and will transmit all the $k_{x}$ components. An imperfect lens will only transmit a finite set of $k_{x}$ components, i.e.$-k_{{\rm max}} \le {\rm k}_{x} \le +k_{{\rm max}} $. The resolution $\Delta$ of such an imperfect lens is then given by \cite{Pendry00}
\begin{equation} \label{GrindEQ__1_}
\Delta =\frac{\pi }{k_{{\rm max}} }
\end{equation}
For TM polarization, the magnetic field of a two-slit object $H_{0y} $ can be written as \cite{Lu03}:
\begin{equation} \label{GrindEQ__2_}
{\rm H}_{0y} =\mathop{\smallint }\nolimits_{0}^{\infty } dk_{x} \upsilon _{k_{x} } e^{ik_{z} z} (\widehat{{\rm Y}}\cos k_{x} x)
\end{equation}
with the coefficient $\upsilon _{k_{x} }$ given by $\upsilon _{k_{x} } =(4/\pi k_{x} )\sin k_{x} a\cos k_{x} b$ and $k_{z} =\sqrt{k_{0}^{2} -k_{x}^{2} } $; $k_{0} =2\pi /\lambda$.

\noindent  Here, 2\textit{b }is the distance between the two slits and 2\textit{a }is the width of the individual slit, as shown in Fig.~\ref{bdfc_fig1}(a). The waves will propagate freely in the vacuum for $k_{x} <k_{0} $. Since $k_{z} $\textit{ }is purely imaginary for evanescent waves with $k_{x} >k_{0} $, the near-field part of $E_{0} $ of the two-slit object is real.

From Eq. \eqref{GrindEQ__2_}, the transmitted near-field of the two-slit object $H_{yt} $ from a flat lens of resolution $\Delta $ reads
\begin{equation} \label{GrindEQ__3_}
{\rm H}_{yt} =\mathop{T_{0} \smallint }\nolimits_{0}^{\pi /\Delta } dk_{x} \upsilon _{k_{x} } e^{ik_{z} z} (\widehat{{\rm Y}}\cos k_{x} x)
\end{equation}

  Here, $T_{0}$ is the transmission coefficient. The imaging performance or resolution of the metamaterials nanolens can be deduced by fitting the transmitted intensity profile curves from the theoretical flat lens, with different values of $\Delta $, to the experimental near-field scanning optical microscope intensity profile. The best fit will represent the actual resolution of the lens. The analytical equations discussed in this letter were computed using MATLAB \cite{matlab07}.

\begin{figure}
  \includegraphics[scale=1]{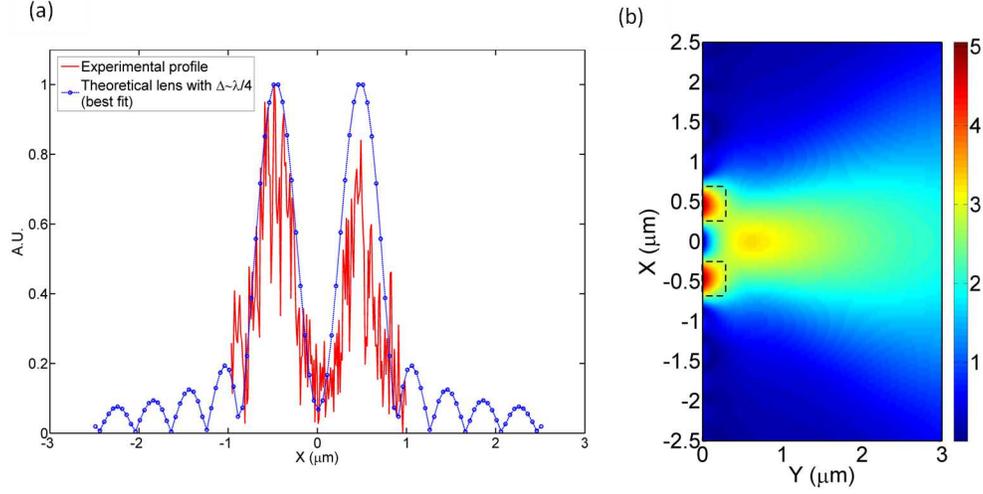}\\
  \caption{\textbf{(a)} Experimental near-field profile of the image of a two-slit object resolved by the 3D metamaterials nanolens of Casse \textit{et al.} \cite{Casse09} (red curve, noisy). The two-slit object comprised of two 600 nm slits spaced 400 nm apart and imaged at 1550 nm wavelength. The best theoretical fit (blue w/ dots) by the analytical method corresponds to a resolution of $\Delta _{theory} \sim \lambda /4$. In this case, this coincides exactly with the experimental estimation. i.e. $\Delta _{\exp } =(d_{min} /\lambda )^{-1} \sim (400/1550)^{-1} \sim \lambda /4$.  This implies that the authors have determined $d_{min}$ correctly. \textbf{(b)} Analytical intensity plot of the two-slit image at the exit (Y=0 $\mu$m) of the theoretical lens. The two dashed lines represent the slits image.}\label{bdfc_fig2}
\end{figure}

 As a first case example, we consider the three-dimensional (3D) metamaterials nanolens by Casse \textit{et al.} in Ref \cite{Casse10}. The two-slit object in this case is composed of two 600 nm slits spaced 400 nm (0.26$\lambda$) and imaged at $\lambda $=1550 nm. The intensity profile of the source object resembles two square-shape pulses, indicating an ideal profile (with a very large resolution).  On the other hand, the diffraction-limited profile has a Gaussian-like shape, where the gap cannot be distinguished.  As for the intensity profile of the image, it has the shape of the letter `M'. For the 3D metamaterials nanolens, the intensity profile curve obtained with a theoretical flat lens with $\sim \lambda /4$ resolution fitted very well the experimental data, as shown in Fig.~\ref{bdfc_fig2}(a). A corresponding analytical intensity plot of the two-slit image at the exit of the theoretical lens is shown in Fig.~\ref{bdfc_fig2}(b). Experimentally, the resolution was found to be \textit{$(d_{min} /\lambda )^{-1} \sim \lambda/4$}.  For this particular case, the authors determined $d_{min} $ accurately. \textit{It is important to note that fitting the two peaks is not as critical as fitting the middle part of the curve}.

\begin{figure}
  \includegraphics[scale=1]{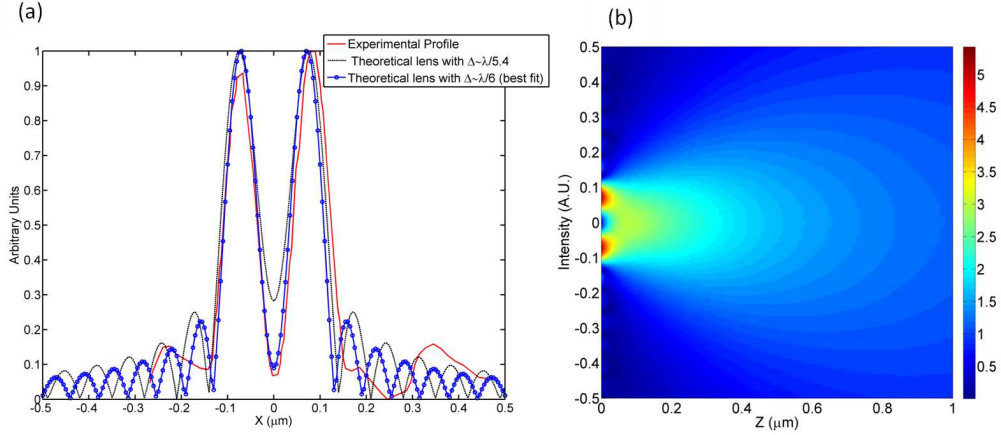}\\
  \caption{\textbf{(a)} Experimental near-field profile of the image of a two-slit object resolved by the far-field superlens in Ref. [16] (red curve). The two-slit object comprised of a nanowire pair with two 50 nm wide slits, spaced 70 nm apart and imaged at 377 nm wavelength. The resolution estimate of the lens is $\Delta _{\exp } =(d_{min} /\lambda )^{-1} \sim (70/377)^{-1} \sim \lambda /5.4$. A curve with $\Delta _{theory} =\lambda /5.4$ (black dashed line) was plotted but did not match the experimental ballpark figure for this case. The best theoretical fit (blue w/ dots) by the analytical method corresponds to a resolution of$\Delta _{theory} \sim \lambda /6$. This indicates that the authors have not reached $d_{min} $ (the size of the gap) yet. Imaging of two slits with a spacing of $\sim $60 nm instead of 70 nm would have been possible. \textbf{(b)} Analytical intensity plot of the two-slit image at the exit (Y=0 $\mu$m) of the theoretical lens.}\label{bdfc_fig3}
  \end{figure}

The next example that we consider is the far-field optical superlens by Z. Liu \textit{et al.} in Ref \cite{Liu07}.  In this case, the object imaged is a nanowire pair with 50 nm wide slits, having a 70 nm gap, at $\lambda $=377 nm, indicating a resolution of $\lambda/5.4$.  The theoretical curve which fitted the experimental data is the theoretical lens that has a resolution of $\lambda /6$, as shown in Fig.~\ref{bdfc_fig3}(a). A corresponding analytical intensity plot of the two-slit image at the exit of the theoretical lens is shown in Fig.~\ref{bdfc_fig3}(b). The curve for a hypothetical lens with $\lambda /5.4$ is also shown for comparison. Our method showed that $d_{min} $ has not been reached in this case, indicating that $d_{min}\sim$ 60 nm instead of 70 nm.

\begin{figure}
  \includegraphics[scale=1]{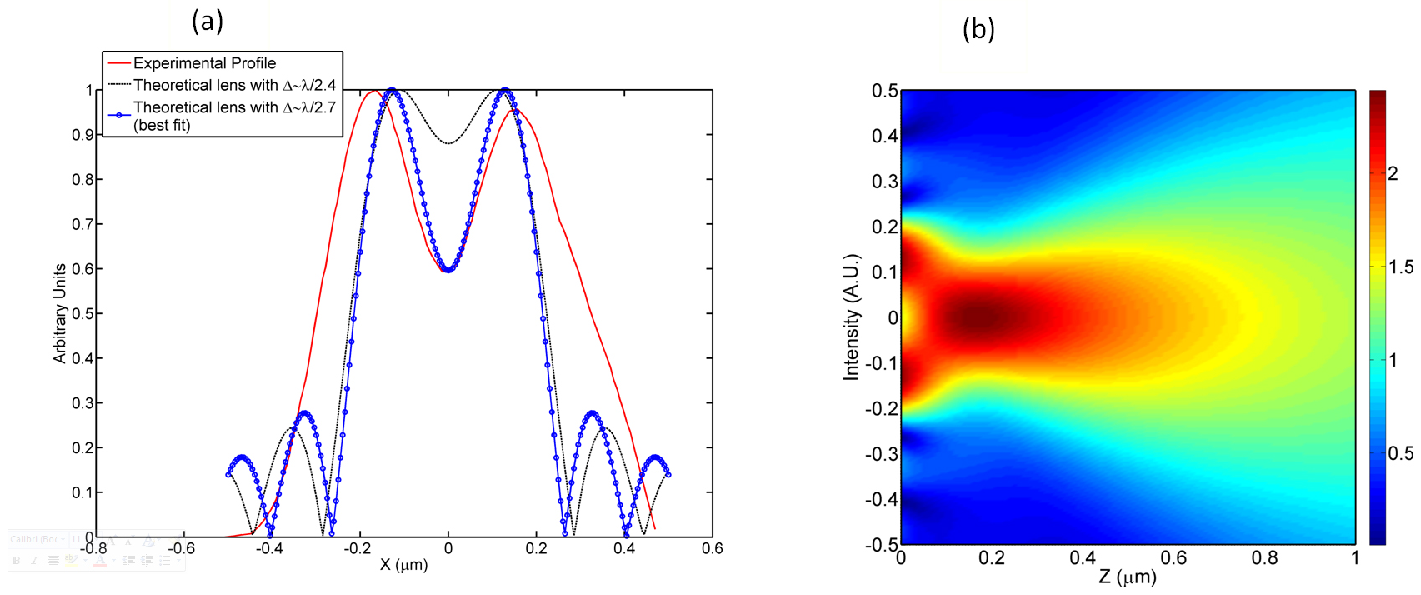}\\
  \caption{\textbf{(a)} Experimental near-field profile of the image of a two-slit object resolved by the far-field optical hyperlens in Ref. [17] (red curve). The two-slit object comprised of a 150-nm-spaced line pair with two 35 nm wide slits imaged at 365 nm wavelength. The resolution estimate of the lens is $\Delta _{\exp } =(d_{min} /\lambda )^{-1} \sim (150/365)^{-1} \sim \lambda /2.4$. A curve with $\Delta _{theory} =\lambda /2.4$ (black dashed line) was plotted but did not match the experimental estimate. The best theoretical fit (blue w/ dots) by the analytical method corresponds to a resolution of $\Delta _{theory} \sim \lambda /2.7$. This indicates that the authors have not reached $d_{min} $ (the size of the gap) yet. Imaging of two slits with a spacing of $\sim $135 nm instead of 150 nm would have been possible. Note that the peaks of the experimental and theoretical curves do not fit because the hyperlens is a magnifying lens. Nevertheless, the magnification doesn't impact the way the resolution is determined and in our method, it is the middle dip of the curve which matters most (i.e. which has to fit). \textbf{(b)} Analytical intensity plot of the two-slit image at the exit (Y=0 $\mu$m) of the theoretical lens.}\label{bdfc_fig4}
\end{figure}

The final example that we discuss is the far-field optical hyperlens by Z. Liu \textit{et al.} in Ref \cite{Liu07_2}. The object imaged at $\lambda $=365 nm was a 150-nm-spaced line pair object with slits of 35 nm width.  The experimental prediction of resolution is $\lambda/2.4$. In this case, the curve which fitted the experimental data is the one with the theoretical lens having a resolution of $\lambda /2.7$, as shown in Fig.~\ref{bdfc_fig4}(a). A corresponding analytical intensity plot of the two-slit image at the exit of the theoretical lens is shown in Fig.~\ref{bdfc_fig4}(b). For reference purposes, the theoretical curve for a lens with resolution of $\lambda /2.4$ is plotted. This suggests that $d_{min} $ is 135 nm instead of 150 nm. Note that the two peaks do not fit exactly simply because of the magnifying property of the hyperlens. Nevertheless, as mentioned above, the magnification doesn't impact the way the resolution is determined, since it's only the middle dip of the curve which matters.

\section{Conclusion}

In summary, we presented an analytical method for determining the resolution of a metamaterials-based superlens by analyzing the transmitted field profile of a two-slit object. The experimental data of the near-field image profile was compared to analytical simulations of the transmitted near-field profile of a hypothetical two-slit object in the vicinity of a theoretical negative-index flat lens. The theoretical flat lens resolution can be varied, with small increments, to fit the experimental data. And, the best fit will reveal the resolution of the superlens. To demonstrate the validity of the method, we have computed the resolution of 3 superresolution lenses from the literature. Moreover, for the last 2 example cases, we showed that the resolution predicted experimentally is slightly lower than the actual resolution of the lens. In other words, imaging would still be possible with slit object gaps reduced by 10--15 nm. The method introduced in this letter is a useful and robust tool for accurately determining the resolution of fabricated superlenses.

\section{Acknowledgments}
The authors would like to thank S. Savo, J. Topolancik and Y. H. Ye for useful discussions and comments. The computations in this paper were run on the Odyssey cluster supported by the Faculty of Arts and Sciences Research Computing Group at Harvard University. This work was financially supported by the Air Force Research Laboratories, Hanscom through FA8718-06-C-0045 and National Science Foundation (NSF) through PHY-0457002.

\newpage

\newpage

\newpage

  \newpage


\begin{thebibliography}{99}
\bibitem{Shalaev07} \textbf{V. M. Shalaev.} Optical negative-index metamaterials. \textit{Nat. Photon.} 1, 41--48 (2007).
\bibitem{Hoffman07} \textbf{A. J. Hoffman \textit{et al.}} Negative refraction in semiconductor metamaterials. \textit{Nat. Mater.}~6, 946--950 (2007).
\bibitem{Lezec07} \textbf{H. J.~Lezec, J. A. Dionne, and H. A.~Atwater.} Negative Refraction at Visible Frequencies. \textit{Science} 316, 430 (2007).
\bibitem{Moser05} \textbf{H. O. Moser, B. D. F. Casse, O. Wilhelmi and B. T. Saw.} \textit{Phys. Rev. Lett.} 94(6):063901 (2005).
\bibitem{Pendry00} \textbf{J. B. Pendry.} Negative Refraction Makes a Perfect Lens. \textit{Phys. Rev. Lett.} 85, 3966--3969 (2000).
\bibitem{Kawata09} \textbf{S. Kawata, Y. Inouye, and P. Verma.} Plasmonics for near-field nano-imaging and superlensing. \textit{Nat. Photon.} 3, 388--394 (2009).
\bibitem{Liu08} \textbf{X . Zhang \& Z. Liu.} Superlenses to overcome the diffraction limit. \textit{Nat. Mater.}~7, 435--441 (2008).
\bibitem{Fang05} \textbf{N. Fang, Hyesog Lee, Cheng Sun, Xiang Zhang.} Sub--Diffraction-Limited Optical Imaging with a Silver Superlens. \textit{Science} 308, 534--537 (2005).
\bibitem{Taubner06} \textbf{T. Taubner, D. Korobkin, Y. Urzhumov, G. Shvets, and R. Hillenbrand.} Near-Field Microscopy Through a SiC Superlens. \textit{Science} 313,~5793 (2006).
\bibitem{Casse09} \textbf{B. D. F. Casse \textit{et al.}} Imaging with subwavelength resolution by a generalized superlens at infrared wavelengths. \textit{Opt. Lett.} 34(13) 1994--1996 (2009).
\bibitem{Smolyaninov07} \textbf{I.  I. Smolyaninov \textit{et al.}} Magnifying Superlens in the Visible Frequency Range. \textit{Science} 315, 1699--1701 (2007).
\bibitem{Merlin07} \textbf{R. Merlin.} Radiationless Electromagnetic Interference: Evanescent-Field Lenses and Perfect Focusing. \textit{Science} 317, 927 (2007).
\bibitem{Lu03} \textbf{W. T. Lu and S. Sridhar.} Near Field Imaging by Negative Permittivity Media. \textit{Microwave and Optical Technology Letters} 39(4 ), 282 (2003).
\bibitem{matlab07} MATLAB version R2010a. Natick, Massachusetts: The MathWorks Inc., 2010.
\bibitem{Casse10} \textbf{B. D. F. Casse \textit{et al.}} Super-Resolution Imaging Using a Three-Dimensional Metamaterials Nanolens. Appl.Phys. Lett., V.96, P.023114 (2010).
\bibitem{Liu07} \textbf{Z. Liu \textit{et al.}} Far-Field Optical Superlens. \textit{Nano Lett.}~(2), 403--408 (2007).
\bibitem{Liu07_2} \textbf{Z. Liu \textit{et al.}} Far-Field Optical Hyperlens Magnifying Sub-Diffraction-Limited Objects. \textit{Science} 315, 1686 (2007).
\end{thebibliography}
\end{document}